\documentclass[preprintnumbersprl,aps]{revtex4}
\usepackage{graphicx}
\usepackage{dcolumn}
\usepackage{bm}
\usepackage{graphics}
\begin{document}
\title{A Family of Exact, Analytical Time Dependent Wave Packet Solutions to a
Nonlinear Schr\"odinger Equation}
\author{S. Curilef$^1$, A.R. Plastino$^{2,3}$ and  A. Plastino$^{4}$ \\
$^1$Departamento de F\'\i sica, Universidad Cat\'olica del Norte, Antofagasta, Chile\\
$^2$CREG-National University La Plata-CONICET, C.C. 727, 1900 La Plata, Argentina \\
$^3$Instituto Carlos I, Universidad de Granada, Granada, Spain \\
$^4$IFLP-National University La Plata-CONICET, C.C. 727, 1900 La Plata, Argentina}
\date{\today}
\begin{abstract}
We obtain time dependent $q$-Gaussian wave-packet solutions to a non linear Schr\"odinger equation
recently advanced by Nobre, Rego-Montero and Tsallis (NRT) [Phys. Rev. Lett. {\bf 106} (2011) 10601].
The NRT non-linear equation admits plane wave-like solutions ($q$-plane waves) compatible with the celebrated
de Broglie relations connecting wave number and frequency, respectively, with energy and momentum.
The NRT equation, inspired in the $q$-generalized thermostatistical formalism, is characterized
by a parameter $q$, and in the limit $q \to 1$ reduces to the standard, linear Schr\"odinger equation.
The $q$-Gaussian solutions to the NRT equation investigated here admit as a particular instance the previously
known $q$-plane wave solutions. The present work thus extends the range of possible processes yielded
by the NRT dynamics that admit an analytical, exact treatment. In the $q \to 1$ limit the $q$-Gaussian
solutions correspond to the Gaussian wave packet solutions to the free particle linear Schr\"odinger
equation. In the present work we also show that there are other families of nonlinear Schr\"odinger-like
equations, besides the NRT one, exhibiting a dynamics compatible with the de Broglie relations.
Remarkably, however, the existence of time dependent Gaussian-like wave packet solutions is a
unique feature of the NRT equation not shared by the aforementioned, more general, families of
nonlinear evolution equations.

\end{abstract}
\pacs{
05.90.+m,
02.30.Jr,
05.45.Yv
}

\keywords{nonlinear Schr\"odinger equation, power laws, wave packets}
\maketitle
\section{Introduction}
A nonlinear Schr\"odinger equation has been recently advanced
by Nobre, Rego-Monteiro and Tsallis  \cite{NRT11,NRT12}.
The NRT proposal constitutes an intriguing contribution to
a line of enquiry that has been the focus of continuous research
activity for several years: the exploration of non-linear
versions of some of the fundamental equations of physics
\cite{S07,SS99}. The NRT equation is inspired in the thermostatistical formalism
based upon the Tsallis $S_q$ non-additive, power-law entropic functional,
whose applications to the study of diverse physical system and processes
have attracted considerable attention in recent years (see, for instance,
\cite{GT04,T09,N11,AdSMNC10,B09} and references therein). In particular,
the $S_q$ entropy constitutes a useful tool for the analysis of diverse
problems in quantum physics \cite{ZP10,SAA10,VPPD12,MML02,ZP06,CAFA11,SSAA12,TBD98}.

 The NRT nonlinear Schr\"odinger equation governing the field
$\Phi(x,t)$ (``wave function'')  corresponding to
a particle of mass $m$ reads \cite{NRT11,NRT12},
\begin{equation} \label{qsnl}
i\hbar \frac{\partial }{\partial t} \left[\frac{\Phi(x,t)}{\Phi_0} \right]=
-\frac{1}{2-q}\frac{\hbar^2}{2m}\frac{\partial^2 }{\partial x^2}
\left[\frac{\Phi(x,t)}{\Phi_0} \right]^{2-q},
\end{equation}
where the scaling constant $\Phi_0$ guarantees an appropriate physical
normalization for the different terms appearing in the equation,
$i$ is the imaginary constant, $\hbar$ is Planck's constant, and
 $q$ is a real parameter formally associated with Tsallis' entropic
index in non-extensive  thermostatistics \cite{T09}. It was shown in
\cite{NRT11} that the wave equation (\ref{qsnl}) admits time dependent
solutions having the ``$q$-plane wave'' form,
\begin{equation} \label{qplane}
\Phi(x,t) \, = \, \Phi_0 \left[1 + (1-q) i (kx - wt) \right]^{\frac{1}{1-q}},
\end{equation}
with $k$ and $w$ real parameters having, respectively, dimensions of
inverse length and inverse time (that is, $k$ can be regarded as a
wave number and $w$ as a frequency). In the limit $q \to 1$, the
$q$-plane waves (\ref{qplane}) reduce to the plane wave solutions
$\Phi_0 \exp(-i(kx-wt))$ of the standard, linear Schr\"odinger
equation describing a free particle of mass $m$.

The $q$-plane wave solutions (\ref{qplane}) propagate at a
constant velocity $c=w/k$ without changing shape, thus exhibiting a
soliton-like behaviour. Moreover, and in contrast to the $q \to 1$
case yielding standard plane waves, the solutions corresponding
to $q \ne 1$ don't have a spatially constant modulus. In fact
(defining $\psi=\Phi/\Phi_0$) we have,
\begin{equation}
|\psi(x,t)|^2 \, = \, \left[1 + (1-q)^2 (kx - wt)^2) \right]^{\frac{1}{1-q}},
\end{equation}
which corresponds, for $1 < q < 3$ to a
normalizable $q$-Gaussian centered at $x=wt/k$.
Therefore, in this case the $q$-plane wave solution describes a
phenomenon characterized by a certain degree of spatial localization.
A field-theoretical approach to the NRT equation was developed
in \cite{NRT12}, where it was shown that this equation can be
derived from a variational principle.
The nonlinear NRT equation is formally related to the nonlinear
Fokker-Planck equation (NLFP) with a diffusion term depending on a power
of the density. These kind of evolution equations, and their
relations with the nonextensive thermostatitical formalism,
have been the focus of an intensive research recently
\cite{F05,RNC12,OW10,TFCP07,AdSdSLL06,PMBML05,FF05,MPP98,U09,TB96}.
In spite of the formal resemblance between the NRT Schr\"odinger
equation and the nonlinear Fokker-Planck, there are profound differences
between these two types of equations. For instance, the nonlinear
Fokker-Planck equation does not admit $q$-plane wave solutions
of the form (\ref{qplane}), that propagate without changing their
shape.

A property of the solutions (\ref{qplane}) that was highlighted by
NRT \cite{NRT11} is that they are consistent with the celebrated
de Broglie relations \cite{P92},
\begin{eqnarray} \label{earlyqu}
E \, = \, \hbar w, \cr
p \, = \, \hbar k,
\end{eqnarray}
connecting, respectively, energy with frequency and
momentum with wave number. Indeed, it can be verified that
the $q$-plane wave (\ref{qplane}) satisfies the equation
(\ref{qsnl}) if and only if the parameters $w$ and $k$ comply
with the relation,
\begin{equation}
w \, = \, \frac{\hbar k^2}{2m},
\end{equation}
which, combined with (\ref{earlyqu}), lead to the
standard relation between linear momentum and kinetic energy,
\begin{equation} \label{enermom}
E \, = \, \frac{p^2}{2m}.
\end{equation}
This suggests that it is conceivable that the $q$-plane wave (\ref{qplane})
represents a particle of mass $m$ with kinetic energy $\hbar w$ and momentum
$\hbar k$ \cite{NRT11}.

 Wave packets (in particular Gaussian wave packets) are of paramount
importance in quantum mechanics, both from the conceptual and practical
points of view \cite{P92,S94}. Wave packets also played a distinguished role
in the historical development of quantum physics \cite{BV09}. It is
interesting to explore the existence of time dependent wave packet solutions
of the NRT nonlinear Schr\"odinger equation. The original presentation of
the NRT equation \cite{NRT11} was strongly focused upon the $q$-plane wave
solutions. However, as a first step towards elucidating the meaning of the NRT
equation, it is imperative to explore more general solutions. The aim of the
present effort is to investigate a family of exact analytical time dependent
solutions to the NRT equation exhibiting the form of $q$-Gaussian wave packets
and corresponding, in the limit $q \to 1$, to the celebrated Gaussian wave
packets solutions of the linear Schr\"odinger equation.


\section{Time Dependent Wave Packet Solutions}

In this work we are going to investigate solutions to the
NRT equation based upon the $q$-Gaussian wave packet ansatz,
\begin{equation} \label{qgauswp}
\psi(x,t) \, = \, \frac{\Phi(x,t)}{\Phi_0} \, = \, \left[1-(1-q)(a(t)x^2+b(t)x+c(t)) \right]^{\frac{1}{1-q}},
\end{equation}
where $a$, $b$, and  $c$ are appropriate (complex) time dependent coefficients.
Notice that $\psi$ depends on time only through these three parameters.
Inserting the ansatz (\ref{qgauswp}) into the left and the right
hand sides of the NRT equation (\ref{qsnl}) one obtains,
\begin{equation}
i\hbar\frac{\partial \psi}{\partial t}=-i\hbar
(\dot{a}(t)x^2+\dot{b}(t)x+\dot{c}(t))\psi^q\label{der_1},
\end{equation}
and,
\begin{equation}
-\frac{1}{2-q}\frac{\partial^2 \psi^{2-q}}{\partial x^2} \, =
\, \left[\frac{}{}-2(3-q)a(t)^2x^2-2(3-q)a(t)b(t)x+2a(t)-
2(1-q)a(t)c(t)-b(t)^2\right]\psi^q \label{der_2}.
\end{equation}
Combining now equations (\ref{qsnl}), (\ref{der_1}), and (\ref{der_2})
one sees that the ansatz (\ref{qgauswp}) constitutes a solution of the
NRT equation provided that the coefficients $a$, $b$, and $c$ comply with the
set of coupled ordinary differential equations,
\begin{eqnarray} \label{qudyn}
i \dot{a}(t) &=&\frac{\hbar}{m}(3-q)a(t)^2\label{coup_10}\\
i \dot{b}(t) &=&\frac{\hbar}{m}(3-q)a(t)b(t)\label{coup_20}\\
i \dot{c}(t) &=&\frac{\hbar}{m}\left((1-q)a(t)c(t)-a(t)+\frac{b(t)^2}{2}\right)\label{coup_30}.
\end{eqnarray}
The above set of differential equations admits the general
solution,
\begin{eqnarray} \label{solv}
a(t) &=&\frac{1}{\frac{(3-q)i\hbar t}{m}+\alpha}\\
b(t)&=&\frac{\beta}{\frac{(3-q)i\hbar t}{m}+\alpha}\\
c(t)&=&{\left(\frac{(3-q)i\hbar}{m}t+\alpha\right)^{-\frac{1-q}{3-q}}}
\left[\frac{\left(\frac{(3-q)i\hbar}{m}t+\alpha \right)^{\frac{1-q}{3-q}}}{1-q}+
\frac{\beta^2}{4}\left(\frac{(3-q)i\hbar}{m}t+
\alpha\right)^{\frac{1-q}{3-q}-1} + \gamma - \frac{1}{1-q} \right],
\end{eqnarray}
where $\alpha$, $\beta$, and $\gamma$ are integrations constants
determined by the initial conditions $a(0)$, $b(0)$, and $c(0)$,
\begin{equation}
\alpha \, = \, \frac{1}{a(0)},
\end{equation}
\begin{equation}
\beta  \, = \, \frac{b(0)}{a(0)},
\end{equation}
\begin{equation}
\gamma \, = \, a(0)^{\frac{q-1}{3-q}}
\left(c(0) - \frac{1}{1-q} - \frac{1}{4} \frac{b(0)^2}{a(0)}\right)
+ \frac{1}{1-q}.
\end{equation}

\subsection{Limit Case $q \to 1$}

Let us now briefly consider the limit $q\to 1$
of the evolving $q$-Gaussian wave packet. In this
case the time dependent parameters $a(t)$, $b(t)$,
and $c(t)$ are,

\begin{eqnarray}
a(t) &=&\frac{1}{\frac{2i\hbar t}{m}+\alpha}\\
{b(t)} &=&\frac{\beta}{\frac{2i\hbar t}{m}+\alpha}\\
c(t) &=&\frac{1}{2}\ln \left(\frac{2i\hbar t}{m}+
\alpha\right)+\frac{1}{4}
\left(\frac{\beta^2}{\frac{2i\hbar t}{m}+\alpha}\right)
+ \gamma,
\end{eqnarray}

\noindent
with $\alpha$, $\beta$ and $\gamma$ integration
constants. The general solution can be written as

\begin{eqnarray}   \label{quno}
\psi&=&\lim_{q\rightarrow 1}\left[1-(1-q)(a(t)x^2+b(t)x+c(t)) \right]^{\frac{1}{1-q}}\cr
&=& \exp\left(-\frac{1}{2}\ln \left(\frac{2i\hbar t}{m}+\alpha \right)-
\frac{(\beta/2+x)^2}{\frac{2i\hbar t}{m}+\alpha} + \gamma\right).
\end{eqnarray}

\noindent
Taking now $\exp(4\gamma)=\frac{2\alpha}{\pi}\exp(-\alpha)$
and defining $ k_0 = i\beta /\alpha$, and $\tan{2\theta}=\frac{2\hbar t}{m\alpha}$,
 (\ref{quno}) can be cast under the guise,

\begin{equation}
\psi=\left(\frac{2\alpha/\pi}{\frac{4\hbar^2t^2}{m^2}+\alpha^2}\right)^{1/4}
\exp\left(-i\frac{}{}\left(\theta+\hbar k_0^2 t/2m \right)\right)
\exp(ik_0x) \exp\left(-\frac{(x-\hbar k_0t/m)^2}{\frac{2i\hbar t}{m}+\alpha}\right),
\end{equation}

\noindent
recovering the well known Gaussian wave packet solution
of the standard linear Schr\"odinger equation.

\subsection{Illustrative Example of Wave Packet Evolution for $q=2$.}

The evolution of the time dependent $q$-Gaussian solution
is illustrated in Figure 1, where the square modulus $|\psi|^2$
of an initially localized solution is depicted against the
(nondimensional) time variable
 ${\bar t} = \left(\frac{\hbar a_0}{m} \right) t$
and spatial coordinate ${\bar x} = \frac{1}{\sqrt{a_0}} x$,
for $q=2$ (here $a_0=|a(0)|$ stands for the modulus of the
intial value of the parameter $a$).
It is interesting that in the nonlinear
($q=2$) case, as time progresses, $|\psi(x,t)|^2$ develops two
peaks that depart from each other. This behaviour exhibits some
qualitative similarity with the evolution of an initially
localized particle in the  tight binding model \cite{CCP11}.
\begin{figure}[hb]
\begin{center}
\vspace{-0.8cm}
\includegraphics[scale=0.4,angle=0]{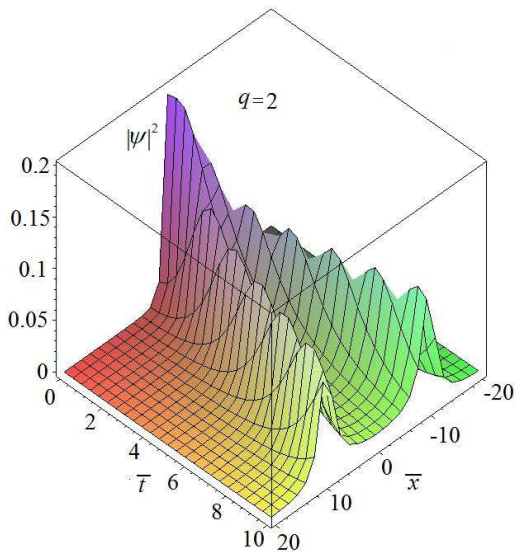}
\vspace{-0.8cm}
\end{center}
\caption{Behaviour of $|\psi(x,t)|^2$ determined by the NRT
nonlinear Schr\"odinger equation with $q=2$ and initial conditions given
by $\alpha=1$, $\beta=1$, and $\gamma=1$. All depiced quantities are
dimensionless.}
\end{figure}


\subsection{The Case $q=3$ and ``Frozen'' Solutions}

When $q=3 $ the set  of coupled differential goberning the evolution
of the parameters $a$, $b$, and $c$, admit the solution,

\begin{eqnarray} \label{abcqutres}
a \, &=& \, a_c, \cr
b \, &=& \, b_c, \cr
c \, &=& \, \frac{b_c^2 - 2a_c}{4 a_c} +
c_1 \exp\left(2i \frac{\hbar}{m} a_c t \right),
\end{eqnarray}

\noindent
where $a_c$, $b_c$, and $c_1$, are time independent constants. The concomitant
solution to the NRT equation reads,

\begin{equation} \label{wqutres}
\psi(x,t) \, =\, \frac{1}{\sqrt{2}} \left[
a_c x^2 + b_c x +
\frac{b_c^2}{4 a_c} +
c_1 \exp\left(2i \frac{\hbar}{m} a_c t\right)
\right]^{-\frac{1}{2}}.
\end{equation}

\noindent
It is interesting that the solution (\ref{wqutres}) is periodic
(with period, in the dimensionless time variable
${\bar t} = \frac{\hbar}{m} a_c t$, equal to $\pi$)
even though the particle is not subjected to an external
confining potential. Indeed, this solution describes
a quasi-stationary scenario where the shape of $|\psi(x,t)|^2$
``pulsates'' with the abovementioned period. In the extreme case
given by $c_1=0$, the amplitude of the ``pulsations'' vanishes
and we obtain the stationary solution,

\begin{equation} \label{qutrequieta}
\psi(x) \, =\, \frac{1}{\sqrt{2}} \left[
a_c x^2 + b_c x +
\frac{b_c^2}{4 a_c}
\right]^{-\frac{1}{2}}.
\end{equation}

\noindent
The norm $N = \int |\psi|^2 dx$ of this last solution
is finite provided that $a_c \ne 0$ and $b_c/a_c$
is not a real number.

It is interesting that there are ``frozen'' solutions
like (\ref{qutrequieta}) also for instances of the
NRT equation characterized by other values of $q$.
Indeed, it can be verified that the NRT equation admits
 stationary solutions of the form

\begin{equation} \label{frozen}
\psi(x) \, =\, [b x + i c]^{\frac{1}{2-q}},
\end{equation}

\noindent
with $b\ne 0$ and $c \ne 0$ real constants.
The square modulus $|\psi|^2$ of the ``frozen''
solutions (\ref{frozen}) has a $q$-Gaussian
profile,

\begin{equation}
|\psi(x)|^2 \, = \, \left[b^2 x^2 + c^2 \right]^{\frac{1}{2-q}}
\end{equation}

\noindent
and a finite norm for $2<q<4$. Notice that the solutions (\ref{frozen})
can be cast under the guise of a $q$-exponential, but with a $q$-value
given by ${\tilde q} = q-1$, which is different from the
$q$-value associated with the concomitant nonlinear NRT
evolution equation. Therefore, strictly speaking, the solutions
(\ref{frozen}) are not comprised within those corresponding to
the ansatz (\ref{qgauswp}).

\subsection{$q$-Plane Waves and Related Particular Solutions}

The case $a=0$ leads to a family of particular
solution of the set of equations (\ref{qudyn}),
given by,

\begin{eqnarray} \label{simple}
a \, &=& \, 0, \cr
b \, &=& \, b_c,  \cr
c \, &=& \, -i \frac{\hbar b_c^2}{2m} \, t \, + c_0
\end{eqnarray}

\noindent
where $b_c$ and $c_0$ are complex constants.
A simple but important case is obtained
when the constant $b_c$ is purely imaginary
and $c_0=0$,

\begin{eqnarray} \label{simpler}
a \, &=& \, 0, \cr
b \, &=& \, -i k, \,\,\, k \in {\mathbf R}, \cr
c \, &=& \, i \frac{\hbar k^2}{2m} \, t,
\end{eqnarray}

\noindent
which, after setting $w= \hbar k^2/2m$, is clearly
seen to correspond to the $q$-plane wave solutions
considered in \cite{NRT11}. Therefore, the exact solutions
to the NRT equation studied in \cite{NRT11} constitute a
particular case of the time dependent $q$-Gaussian wave
packet investigated here.

Another interesting particular instance of the solutions
associated with (\ref{simple}) corresponds to the case
where the constant $b_c$ is a real number and
$c_0 =  -i \frac{\hbar b_c^2}{2m} t_0$, with $t_0>0$ a real
constant with dimensions of time, leading to,

\begin{equation} \label{simplunexpected}
\psi(x,t) \, = \, \left[1 + (1-q) \left(-b_c x +
\frac{i \hbar}{2m} b_c^2 (t+t_0) \right) \right]^{\frac{1}{1-q}}.
\end{equation}

\noindent
The squared modulus of the above soution has
a $q$-Gaussian form,

\begin{equation} \label{simplenorma}
|\psi(x,t)|^2 \, = \, \left[\frac{ (1-q)^2 \hbar^2}{4m^2}
b_c^4 (t+t_0)^2 \right]^{\frac{1}{1-q}}
\left\{ 1 +
\left[
\frac{2m(1-(1-q)b_c x)}{(1-q)\hbar b_c^2 (t+t_0)}
\right]^2
\right\}^{\frac{1}{1-q}},
\end{equation}
leading to a finite norm, $N = \int |\psi|^2 dx$, for
$t>-t_0$ and $1<q<3$ (see next subsection).
The solution (\ref{simplunexpected})
exhibits a finite-time singularity in the past:
indeed, its norm diverges at the finite time
$t=-t_0$. In the limit $q \to 1$ the solutions
(\ref{simplunexpected}) go to

\begin{equation} \label{expoexplodes}
\psi(x,t) \, = \, \exp(-b_c x) \exp \left[
\frac{i \hbar}{2m} b_c^2 (t+t_0) \right],
\end{equation}

\noindent
which are formal solutions of the
standard linear Schr\"odinger equation but are,
evidently, physically unacceptable because
they are not normalizable (and the wave function's
square modulus $|\psi|^2$ itself diverges
when $x b_c \to -\infty$). It is
nevertheless interesting that the non-linearity
associated with $q>1$ not only regularizes
(in the sense of leading to a finite norm $N$)
the plane wave solutions of the Schr\"odinger
equation (as stressed by NRT in \cite{NRT11})
but it also regularizes the exponential
solutions (\ref{expoexplodes})
(at least for all times $t> -t_0$).

\subsection{(Non-)Preservation of the Norm}

It is known that, in general, time dependent solutions to the NRT nonlinear Schr\"odinger
equation do not preserve the norm \cite{NRT12}. The $q$-plane wave solutions studied in
\cite{NRT12} constitute a remarkable exception: they do preserve the norm. Up
to now the $q$-plane wave solutions where the only known time dependent solutions
to the NRT equation. This means that no norm non-preserving explicit solution
was known before our present work. Therefore, it is of some interest to explore
whether the solutions to the NRT equation investigated here preserve the norm or not.
The norm $N$ of the $q$-Gaussian wave packet is given by,
\begin{equation}
N=\int_{-\infty}^\infty dx|\psi|^2,
\end{equation}
where
\begin{equation}
|\psi|^2=\left[1-2(1-q)\Re(a(t)x^2+b(t)x+c(t))+
(1-q)^2|a(t)x^2+b(t)x+c(t)|^2\right]^{\frac{1}{1-q}}.
\end{equation}
When $a\!\! \ne \!\!0 $ the $q$-Gaussian wave packet is normalizable (that is, $N < \infty$)
provided that  $1 < q < 5$ and the polynomial $P(z) = 1-(1-q)(a(t)z^2+b(t)z+c(t))$ doesn't
have real roots. Notice that for $a \ne 0$ the range of $q$-values
admitting normalizable $q$-Gaussian wave functions of the form
(\ref{qgauswp}) is larger than the range of $q$-values leading
to normalizable $q$-plane wave functions (which is $1 < q < 3$
\cite{NRT11}).

Let us now consider a concrete instance of a solution to the NRT
equation that does not preserve the norm.
In the case of the solution (\ref{simplunexpected}) the norm is,

\begin{eqnarray}
N \, &=& \, \frac{1}{|1-q| b_c} \left[\frac{(1-q)^2 \hbar^2}{4 m^2}
b_c^4 (t+t_0)^2 \right]^{\frac{1}{2} + \frac{1}{1-q}}
\int_{-\infty}^{+ \infty} du \left[1 + u^2 \right]^{\frac{1}{1-q}} \cr
\, &=& \, \frac{1}{|1-q| b_c} \left[\frac{(1-q)^2 \hbar^2}{4 m^2}
b_c^4 (t+t_0)^2 \right]^{\frac{1}{2} + \frac{1}{1-q}}
\sqrt{\pi } \,\, \frac{\Gamma  \left( {\frac {3-q}{2(q-1)}} \right)}{ \Gamma
 \left(  \frac{1}{q-1} \right)}.
\end{eqnarray}
The norm is finite for $t>-t_0$ and $1<q<3$ and
proportional to $(t+t_0)^{1+ \frac{2}{1-q}}$. Therefore, the
norm is not conserved: it vanishes in the limit $t \to +\infty$,
it is a finite and monotonously decreasing function of time for
all finite times $t>t_0$, and it diverges at $t=-t_0$. Consequently,
(\ref{simplunexpected}) constitutes an explicit example of a norm
non-preserving solution to the NRT equation.

As a final comment, let us mention that the non conservation
of the norm suggests that, when considering pairs
of time dependent solutions $\psi_1$ and $\psi_2 $
to the NRT equation, the overlap $\int \psi_1^{*}\psi_2 dx$
is not conserved either. However, there may still be
some conserved measure of the ``distance'' or ``fidelity''
between pairs of time dependent solutions. The search for
such a measure constitutes a line of enquiry that may shed
new light on the nature of the NRT dynamics. A possible
direction to explore in this regard would be the one
suggested by Yamano and Iguchi in \cite{YI08}, where
non-Csiszar $f$-divergence measures where considered in
connection with nonlinear Liouville-type equations.

\section{Wave Packet Solutions in the Presence of an Harmonic Confining Potential}

 Now we are going to use the $q$-Gaussian ansatz (\ref{qgauswp}) to investigate
time dependent solutions of nonlinear Schr\"odinger equation given by

\begin{equation} \label{quharmo}
i\hbar \frac{\partial \psi}{\partial t}=
-\frac{1}{2-q}\frac{\hbar^2}{2m}\frac{\partial^2 \psi^{2-q}}{\partial x^2}+V(x)\psi^q,
\end{equation}

\noindent
where $V(x)=\frac{1}{2}Kx^2$. In the limit $q\to 1$ the above equation
reduces to the time dependent Schr\"odinger equation of the quantum harmonic
oscillator. Inserting the ansatz (\ref{qgauswp}) in the right hand side
of equation (\ref{quharmo}) we get

\begin{eqnarray}
-\frac{1}{2-q}\frac{\hbar^2}{2m}\frac{\partial^2 \psi^{2-q}}{\partial x^2}+V(x)\psi^q &=&\frac{\hbar^2}{2m}\left[\frac{}{}2a(t)(1-(1-q)(a(t)x^2+b(t)x+c(t)))-(2a(t)x+b(t))^2\right]\psi^q+\frac{1}{2}Kx^2\psi^q\nonumber\\
&=&\frac{\hbar^2}{2m}\left[\left(\frac{m}{\hbar^2}K-2(3-q)a(t)^2\right)x^2-2(3-q)a(t)b(t)x+2a(t)-2(1-q)a(t)c(t)-b(t)^2\right]\psi^q \nonumber
\end{eqnarray}

\noindent
Comparing Eq.(\ref{der_1}) with (\ref{der_2}) we obtain the following
set of coupled differential equations for the evolution of the parameters
appearing in the ansatz

\begin{eqnarray} \label{abc_harmon}
i \dot{a}(t) &=&\frac{\hbar}{m}(3-q)a(t)^2-\frac{K}{2\hbar}\label{coup_1}\cr
i \dot{b}(t) &=&\frac{\hbar}{m}(3-q)a(t)b(t)\label{coup_2}\cr
i \dot{c}(t) &=&\frac{\hbar}{m}\left((1-q)a(t)c(t)-a(t)+\frac{b(t)^2}{2}\right)\label{coup_3}.
\end{eqnarray}

\noindent
Integrating the first of these equation one obtains,

\begin{equation}
\frac{1}{\frac{\hbar}{m}(3-q)}\int \frac{da(t)}{\frac{mK}{2\hbar^2(3-q)}-a(t)^2}
=\frac{1}{\frac{\hbar}{m}(3-q)}\frac{1}{2\sqrt{\frac{mK}{2\hbar^2(3-q)}}}\log\left|
\frac{\sqrt{\frac{mK}{2\hbar^2(3-q)}}+a(t)}{\sqrt{\frac{mK}{2\hbar^2(3-q)}}-a(t)}\right|
=it+\delta,
 \end{equation}

\noindent
where $\delta$ is an integration constant.

\subsection{Quasi-Stationary Solutions}

A particularly interesting solution of the set of equations (\ref{abc_harmon})
is given by,

\begin{eqnarray}
a \, &=& \, a_c \, = \, \frac{1}{\hbar} \sqrt{\frac{mK}{2 (3-q)}}, \cr
b \, &=& \, 0, \cr
c \, &=& \, \frac{1}{1-q} \left[1-  \exp\left(-i(1-q) \frac{\hbar a_c t}{m} \right) \right],
\end{eqnarray}

\noindent
leading, in turn, to the following solution for the
non-linear Schr\"odinger equation,

\begin{equation} \label{quasiest}
\psi \, = \, \left[\exp\left(-i(1-q) \frac{\hbar a_c t}{m} \right)
- (1-q) a_c x^2 \right]^{\frac{1}{1-q}}.
\end{equation}

\noindent
It is interesting to consider now the $q\to 1$ limit of the above
solution,

\begin{eqnarray} \label{qunoest}
\lim_{q \to 1} \psi \, &=& \,
\exp\left(-i\frac{\hbar a_c t}{m} \right)
\lim_{q \to 1}
\left[ 1- (1-q) a_c\exp\left(i(1-q) \frac{\hbar a_c t}{m} \right)
 x^2 \right]^{\frac{1}{1-q}} \cr
\, &=& \, \exp\left(-i\frac{\hbar a_c t}{m} \right)
\exp\left(-a_c x^2 \right) \cr
\, &=& \, \exp\left(-i \frac{\omega t}{2} \right)
\exp\left(- \frac{m \omega}{ 2 \hbar} x^2 \right),
\end{eqnarray}

\noindent
which is the (unnormalized) wave function
associated with the ground state of a
standard harmonic oscillator with natural frequency
$\omega = \sqrt{\frac{K}{m}}$ and zero point energy
$E_0 = \frac{1}{2} \hbar \omega$.

The norm of the solution (\ref{quasiest}) is finite for
$1<q<5$ and $-t_c< t < t_c $ with $t_c = \frac{\pi m}{(q-1)
\hbar a_c}$, and it has finite-time sigularities at
$t= \pm t_c$. The time derivative of the norm of
the solution (\ref{quasiest}) is,

\begin{equation}
\frac{dN}{dt} \, = \,  2 (1-q) \frac{\hbar a_c^2}{m}
\sin \left[ (1-q) \frac{\hbar a_c t}{m} \right]
\int_{-\infty}^\infty
 x^2
\left[1 - 2 a_c (1-q)
\cos \left[ (1-q) \frac{\hbar a_c t}{m} \right]  x^2
+ (1-q)^2 a_c^2 x^4 \right]^{\frac{q}{1-q}} dx.
\end{equation}

\noindent
The expression between square brackets appearing
in the integrand in the right hand side of the
above equation is in general larger than zero and,
consequently, the time derivative of the norm is
different from zero. Therefore, the time dependent
solution (\ref{quasiest}) constitutes another explicit
example of a solution that does not preserve the norm.
In the limit $q \to 1$ we get $|t_c| \to \infty$ and,
of course, $dN/dt \to 0$.

\section{Generalizations of the NRT Approach and
Unique Features of the NRT Equation}

The existence of $q$-plane wave solutions consistent
with the de Broglie relations connecting frequency
and wave number respectively with energy and momentum
was one of the features of NRT equation discussed in
\cite{NRT11}. It is worth noticing that there are other
possible nonlinear Schr\"odinger-like equations exhibiting
similar properties. That is, the NRT approach can be
substantially generalized. Let us consider a pair of one-variable
functions $L(u)$ and $F(u)$ satisfying the functional relation,

\begin{equation} \label{nicecouple}
\frac{d^2}{du^2} \left[ L(F(u)) \right] \, = \, \frac{dF}{du}.
\end{equation}

\noindent
It can then be verified after some algebra that the (in general non linear)
Schr\"odinger -like equation

\begin{equation} \label{genert}
i \hbar \frac{\partial}{\partial t} \left[\frac{\Phi(x,t)}{\Phi_0} \right] \, = \, -\frac{\hbar^2}{2m}
\frac{\partial^2}{\partial x^2}\left[L \left( \frac{\Phi(x,t)}{\Phi_0} \right)\right]
\end{equation}

\noindent
admits exact time dependent plane wave-like solutions of the form,

\begin{equation} \label{genplanwe}
\Phi(x,t) \, = \, \Phi_0 F[i (kx-wt)],
\end{equation}

\noindent with $\hbar w = \hbar^2 k^2 /2m$. Indeed, inserting
the ansatz (\ref{genplanwe}) into the right and left hand sides
(\ref{genert}) and  setting $u= i (kx-wt)$ we get,

\begin{eqnarray}
i \hbar \frac{\partial}{\partial t} \left[\frac{\Phi(x,t)}{\Phi_0} \right]  &=&
\left( \frac{dF}{du} \right) \left( \frac{\partial u}{\partial t} \right)  =
\hbar w \, \frac{dF}{du},
\cr
-\frac{\hbar^2}{2m}
\frac{\partial^2}{\partial x^2}\left[L \left( \frac{\Phi(x,t)}{\Phi_0} \right)\right] &=&
-\frac{\hbar^2}{2m}
\frac{d^2}{d u^2}\left[L \left(u \right)\right] \left(\frac{\partial u}{\partial x} \right)^{\!\!2}
= \frac{\hbar^2 k^2}{2m} \, \frac{d^2}{du^2} \left[ L(F(u)) \right],
\end{eqnarray}

\noindent
where, in the last equation, the fact that $\partial^2 u /\partial x^2 = 0$
was used. It is plain that the functional relation (\ref{nicecouple}) implies
(\ref{genert}), provided that $\hbar w = \hbar^2 k^2 /2m$. This in turn
leads, via the de Broglie relations, to the kinetic energy-momentum
relation $E = p^2/2m$. Similarly to what happens
within the NRT scenario, the solutions (\ref{genplanwe})
to the non linear evolution equation (\ref{genert}) propagate
without changing shape and with constant velocity $v = w/k$. Furthermore,
as we have seen, they comply with a relation between $w$ and $k$ that
is consistent with the de Broglie connection between frequency,
wave number, energy, and momentum.

One procedure to generate pairs of functions satisfying (\ref{nicecouple})
is the following. One starts with a function $G(u)$ such that its derivative
$dG/du$ admits an inverse. Then we define,

\begin{eqnarray} \label{nicepair}
F(u) \, &=& \, \frac{dG}{du}, \cr
L(u) \, &=& \, G\left[ F^{(-1)}(u) \right],
\end{eqnarray}

\noindent
where $F^{(-1)}(u)$ is the inverse function of $F(u)$, satisfing
$F^{(-1)}(F(u))=u$. It can be verified that the functions defined
by (\ref{nicepair}) comply with the required relation
(\ref{nicecouple}).

In the case of the NRT equation we have,

\begin{eqnarray}
G(u) \, &=& \, \frac{1}{2-q} [1+(1-q)u]^{\frac{2-q}{1-q}}, \cr
F(u) \, &=& \, [1+(1-q)u]^{\frac{1}{1-q}}, \cr
L(u) \, &=&  \, \frac{u^{2-q}}{2-q}.
\end{eqnarray}

\noindent
An example (different from the NRT one) of a nonlinear Schr\"odinger -like
equation of the form (\ref{genert}) admitting the plane wave-like solutions
(\ref{genplanwe}) is given by $F(u) = \sinh(u)$ and $L(u) = \sqrt{1+u^2}$.
This case corresponds to the nonlinear evolution equation,

\begin{equation}
i \hbar \frac{\partial}{\partial t} \left[\frac{\Phi(x,t)}{\Phi_0} \right]
\, = \, -\frac{\hbar^2}{2m} \frac{\partial^2}{\partial x^2}
\left\{ \! 1+ \left[\frac{\Phi(x,t)}{\Phi_0} \right]^2 \! \right\}^{ \! \frac{1}{2}},
\end{equation}

\noindent
having plane wave-like solutions

\begin{equation}
\Phi(x,t) \, = \, \Phi_0 \sinh[i (kx-wt)],
\end{equation}

\noindent
with $\hbar w = \hbar^2 k^2 /2m$.

As mentioned in the Introduction, the NRT equation was inspired by
the nonextensive generalized thermostatistical formalism based on
the constrained optimization of Tsallis' entropy \cite{NRT11}.
Indeed, there are formal connections between the NRT scenario
and the alluded thermostatistical formalism. An intriguing
question, which is beyond the scope of the present work but
certainly deserves to be explored, is the possible existence
of connections between (some of) the nonlinear Schr\"odinger
equations (\ref{genert}), on the one hand, and other formalisms
based on non-standard entropic functionals different from the
Tsallis one \cite{B09,Y00}, on the other hand.

We have seen that the NRT equation is not the only nonlinear equation
of the form (\ref{genert}) admitting plane wave-like solutions consistent
with the de Broglie relations. It is then natural to ask which
of the equations (\ref{genert}) also admit solutions of the form

\begin{equation} \label{genagus}
\Phi(x,t) \, = \, \Phi_0 F[d_2(t)x^2 + d_1(t)x + d_0(t)],
\end{equation}

\noindent
with $d_1(t), d_2(t), d_0(t)$ appropriate time dependent
coefficients. The solutions (\ref{genagus}) would constitute
generalizations of the $q$-Gaussian solutions to the NRT
equation previously discussed in the present work. It turns
out that the NRT equation is the only member of the family
(\ref{genert}) admitting both solutions of the form
(\ref{genplanwe}) and of the form (\ref{genagus}).
If one inserts in the left and rigth hand sides of the
evolution equation (\ref{genert}) the expression for
$\psi = \frac{\Phi}{\Phi_0}$ given by the ansatz ansatz
(\ref{genagus}) one gets,

\begin{equation}
i\hbar\frac{\partial \psi}{\partial t}\, = \, i\hbar
(\dot{d_2} x^2+\dot{d_1} x+ \dot{d_0}) F^{\prime}(u),
\end{equation}

\noindent
and,

\begin{equation}
-\frac{\hbar^2}{2m}\frac{\partial^2}{\partial x^2}
\Bigl[L(F(u))]\Bigr]\, = \, -\frac{\hbar^2}{2m}
\Bigl\{(2d_2 x + d_1)^2 \frac{d^2}{du^2} \Bigl[ L(F(u)) \Bigr]
+ 2d_2 \frac{d}{du} \Bigl[ L(F(u)) \Bigr]
\Bigr\},
\end{equation}

\noindent
where $u= d_2(t)x^2 + d_1(t)x + d_0(t)$. Then, to have solutions
of the form (\ref{genplanwe}) (plane wave-like), equation (\ref{nicecouple})
must hold. On the other hand, to have solutions of the form
(\ref{genagus}) one further condition is required,

\begin{equation} \label{errecond}
\frac{d}{du} \Bigl[ L(F(u)) \Bigr] \, = \, (r_1 + r_2 u)  F^{\prime}(u),
\end{equation}

\noindent
with $r_1$ and $r_2$ appropriate constants. It follows from (\ref{errecond}) that,

\begin{equation} \label{erre2}
L^{\prime \prime}(F(u)) F^{\prime} (u) \, = \, r_2.
\end{equation}

\noindent
Combining now equations (\ref{nicecouple}), (\ref{errecond})
 and (\ref{erre2}) one obtains,

\begin{equation} \label{erre12}
r_2  F^{\prime} (u) + (r_1 + r_2 u) F^{\prime \prime}(u) \, = \,  F^{\prime} (u),
\end{equation}

\noindent
which leads to,

\begin{equation} \label{close_to_qexp}
F^{\prime}(u) \, = \, F_0 (r_1 + r_2 u)^{\frac{1-r_2}{r_2}},
\end{equation}

\noindent
where $F_0$ is an integration constant. Finally, we have,

\begin{equation} \label{voila_almost_qexp}
F(u) \, = \, F_0 (r_1 + r_2 u)^{\frac{1}{r_2}} + F_1,
\end{equation}

\noindent
involving one more integration constant $F_1$. Making now the identification
$r_2 = 1-q$ one sees that (up to multiplicative and additive constants)
the form of the solution (\ref{voila_almost_qexp}) coincides with
$q$-Gaussian wave packet.

\section{Conclusions}

We have obtained a new family of exact, analytical
time dependent wave packet solutions to the nonlinear
Schr\"odinger equation recently proposed by
Nobre, Rego-Montero and Tsallis \cite{NRT11,NRT12}.
Our solutions have the form of a $q$-exponential evaluated
upon a quadratic function of the spatial coordinate $x$
with time dependent coefficients. Therefore, these
solutions have a $q$-Gaussian form. They extend and
generalize the previously known solutions to the
NRT equation. The solutions investigated here by us
correspond, in the limit $q \to 1$ of the parameter $q$,
to the Gaussian wave packet solutions to the standard
linear Schr\"odinger equation. Our present wave packet
solutions admit as a special particular case the $q$-plane
wave solutions studied in \cite{NRT11}. In the present
work we also discuss other families of nonlinear
Schr\"odinger-like equations, besides the NRT one,
leading to a dynamics compatible with the de Broglie relations.
In this regard, it is remarkable that the existence of the
Gaussian-like time dependent solutions investigated in
this work is a unique feature of the NRT equation not
shared by the abovementioned, more general, families of
nonlinear evolution equations.

We also obtained $q$-Gaussian wave packet solutions for
the case of a harmonic potential. In the limit $q \to 1$
these latter solutions reduce to Schr\"odinger's celebrated
Gaussian wave packet solutions to the harmonic oscillator.
As a particular case of the time dependent $q$-Gaussian wave packets
associated with the harmonic potential we found a quasi-stationary
solution yielding in the $q \to 1$ limit the wave function corresponding
to the ground state of the quantum harmonic oscillator.

\acknowledgments This work was partially supported by the Projects
FQM-2445 and FQM-207 of the Junta de Andalucia and the grant
FIS2011-24540 of the Ministerio de Innovaci\'on y Ciencia (Spain).


\end{document}